\documentclass[preprint,12pt, number]{elsarticle}

\usepackage{amssymb}
\usepackage{siunitx}
\usepackage{amsmath}
\usepackage{multirow}
\biboptions{numbers,sort&compress}
\usepackage{hyperref}

\usepackage{lineno}
\usepackage{float}
\usepackage{booktabs}
\journal{Nuclear Instruments and Methods A}

\begin{document}

\begin{frontmatter}



\title{Measurements of Rayleigh Ratios in Linear Alkylbenzene}

\author[label1]{Miao Yu}
\author[label1]{Wenjie Wu}
\author[label2,label3]{Na Peng}
\author[label2]{Taozhe Yu}
\author[label2]{Yayun Ding}
\author[label3]{Qian Liu}
\author[label1]{Feng Ren\corref{aaa}} 
\ead{fren@whu.edu.cn}
\author[label1]{Zhenyu Zhang}
\author[label1]{Xiang Zhou\corref{bbb}}
\ead{xiangzhou@whu.edu.cn}
\affiliation[label1]{organization={Hubei Nuclear Solid Physics Key Laboratory, School of Physics and Technology},
           addressline={Wuhan University}, 
            city={Wuhan},
            postcode={430072}, 
            state={Hubei},
            country={China}}
\affiliation[label2]{organization={Institute of High Energy Physics},
           addressline={Chinese Academy of Sciences}, 
            city={Beijing},
            postcode={100049}, 
            country={China}}
\affiliation[label3]{organization={School of Physics},
           addressline={University of Chinese Academy of Sciences}, 
            city={Beijing},
            postcode={100049}, 
            country={China}}
\cortext[aaa]{Corresponding Author}
\cortext[bbb]{Corresponding Author}

\begin{abstract}

In present work an experiment has been designed to measure Rayleigh ratios directly at 405 nm and $432$ nm for linear alkylbenzene which is a common solvent used in liquid scintillator detectors of neutrino experiments. The perpendicularly polarized Rayleigh ratio is determined to be $(4.52\pm 0.28)\times 10^{-6}$ m$^{-1}\cdot$ sr$^{-1}$  at $405$ nm and $(3.82\pm 0.24)\times 10^{-6}$ m$^{-1}\cdot$ sr$^{-1}$  at $432$ nm, and the corresponding Rayleigh scattering length is $L_{Ray} = 22.9\pm 0.3(\mathrm{stat.})\pm 1.7(\mathrm{sys.})$ m at $405$ nm and $L_{Ray}= 27.0\pm 0.9(\mathrm{stat.})\pm 1.8(\mathrm{sys.})$ m at $432$ nm.

\end{abstract}




\end{frontmatter}


\section{Introduction} \label{Sec:Intro}

In recent years, neutrino experiments have entered an era of precision measurement, where liquid scintillator (LS) detectors have played important roles due to many advantages such as high light yield, low radioactive impurities and constructive convenience. For example, the reactor neutrino experiments for precise oscillation parameters measurements in KamLAND~\cite{KamLAND:2008dgz}, Daya Bay~\cite{DayaBay:2012fng}, RENO~\cite{RENO:2012mkc} and Double Chooz~\cite{DoubleChooz:2011ymz}, solar neutrino studies in Borexino~\cite{Borexino:2017rsf} and neutrinoless double beta ($0\nu\beta\beta$) decay exploration in KamLAND-Zen~\cite{KamLAND-Zen:2012mmx} have all benefited from the advantages of LS detectors. The common recipe of LS consists of solvent and fluor, sometimes also the wavelength shifter. Energy deposited by charged particles ionizes and excites the solvent molecules firstly and then non-radioactively transferred to the fluor molecules which generate photons by deexcitation. Due to no concerns on the environment and human health, linear alkylbenzene (LAB) is a popular solvent as in RENO, Daya Bay, SNO+~\cite{SNO:2021xpa} and the upcoming experiment JUNO~\cite{JUNO:2015zny}.

High energy resolution is a necessary indicator for precision measurement experiments. For example, precise determination of neutrino mass ordering in JUNO requires an unprecedented energy resolution of $3\%/\sqrt{E\left[ \mathrm{MeV} \right]}$~\cite{Zhan:2008id,Zhan:2009rs}, 
and high energy resolution is also very beneficial for spectral fits in $0\nu\beta\beta$ decay experiments like SNO+~\cite{SNO:2021xpa}. Due to the extremely weak interactions between neutrinos and the detector materials, LS detectors tend to have large volumes. Photons generated at the bulk have to cross several meters until detection. The absorption of photons in the LS can reduce the transparency and be detrimental to the energy resolution. However, the absorption length of LAB is difficult to measure directly. An indirect calculation method based on the attenuation length and the Rayleigh scattering length has been proposed by Ref.~\cite{Bohren1983}. The total light attenuation in LS detectors results from absorption and Rayleigh scattering, and it can be described by Eq.~\eqref{Eq:Latt},
\begin{equation}
    \frac{1}{L_\mathrm{att}} = \frac{1}{L_\mathrm{abs}} + \frac{1}{L_\mathrm{Ray}}, 
    \label{Eq:Latt}
\end{equation}
where $L_\mathrm{att}$, $L_\mathrm{abs}$ and $L_\mathrm{Ray}$ refer to the attenuation length, absorption length and Rayleigh scattering length respectively. Besides, the Rayleigh scattering length itself is important for more comprehensive understanding of light propagation in large LS detectors. The space and time patterns of photons are modified due to Rayleigh scattering, which requires correct consideration in event reconstruction~\cite{Wu:2018zwk}. Optical processes in the LS detectors are dominated by the solvent due to its large amount, thus the optical properties of LAB are key indices for these experiments. Therefore, precise measurement methods of the Rayleigh scattering length of LAB can have extensive applications in LS experiments. 

Prior to this work, broad measurements have been performed to understand the optical properties of LAB. The current measurement results of $L_\mathrm{att}$ for LAB can be found in Ref.~\cite{Ding:2008zzb,Goett:2011zz,Gao:2013pua,Wurm:2010ad}. The Rayleigh scattering length of LAB has been firstly measured based on a multi-angular experimental setup~\cite{Wurm:2010ad}. However the cylindrical cuvette can distort the facula and has rather complicated surface optical processes which make the suppression of stray light difficult. The Rayleigh scattering length has also been obtained according to the Einstein-Smoluchowski-Cabannes (ESC) formula~\cite{ESC} by measuring the depolarization ratio of LAB in Ref.~\cite{Liu2015,Zhou2015,Zhou:2015fwa}. However, measurements of other LAB properties in the ESC formula, like isothermal compressibility, are time consuming and expensive. Moreover, the ESC formula itself suffers from various expressions which introduces further uncertainties.  

A single angular direct measurement of scattered light intensity in Ref.~\cite{Finnigan1970} motivates the development of a new method for the direct measurement of the Rayleigh scattering length. The model independent method is supposed to be convenient and fast for multi-wavelength measurements based on a modularized experimental setup. In this paper, we measured the Rayleigh scattering length for LAB. A brief introduction of Rayleigh scattering is presented in Section \ref{Sec:Ray}. The experimental setup and calibration strategies are described in Section \ref{Sec:Exp}. Systematic uncertainties are discussed in Section \ref{Sec:Uncer}. And results are given in Section \ref{Sec:Results}.

\section{Rayleigh scattering in LAB}\label{Sec:Ray}
Rayleigh scattering is a kind of radiation by particles much smaller than radiation wavelength. It is well-known that Rayleigh scattering by isotropic molecules is polarized~\cite{Rayleigh1899}. However, for anisotropic liquids like LAB, the polarizability is a tensor property, which induces depolarized components in the scattered light. The depolarization ratio for materials is defined as:
\begin{equation}
	\delta \equiv \frac{I^h(90^\circ)}{I^v(90^\circ)}, \label{Eq:Delta}
\end{equation}
where $I^v(90^\circ)$ and $I^h(90^\circ)$ refer to the intensity of scattered light polarized perpendicular ($v$) and parallel ($h$) to the scattering plane with unpolarized incident light. The appearance of depolarized part in scattered light will shorten the Rayleigh scattering length of LAB as discussions in Ref.~\cite{Zhou2015,Zhou:2015fwa}.

The angular distribution for Rayleigh scattering of anisotropic molecules obeys $(1+(1-\delta)/(1+\delta)\cos^2\theta)$~\cite{Dawson1941}. According to the definite angular distribution, measurements of the Rayleigh scattering length can be 
simplified at one single scattering direction instead of the integrated measurements. The optimal choice is $90^\circ$ for good machinability of cubic cuvette and easy handling of stray light. To characterize Rayleigh scattering intensity at $90^\circ$, the Rayleigh ratio is introduced as: 
\begin{equation}
	R \equiv \frac{I(90^\circ)}{I_0} \frac{r^2}{V}, \label{Eq:RayRatio}
\end{equation}
where $I(90^\circ)$ is the scattered radiation intensity at $90^\circ$ direction, $I_0$ is the incident radiation intensity, $r$ is the distance from the scattering center to the detector and $V$ is the scattering element volume. And the Rayleigh scattering length is expressed as Eq.~\eqref{Eq:LRay} by $4\pi$ solid angle integral,
\begin{equation}
	L_\mathrm{Ray} = \left[ \frac{8\pi}{3}R\frac{2+\delta}{1+\delta}\right]^{-1}. \label{Eq:LRay}
\end{equation}

The calculation in Eq.~\eqref{Eq:LRay} is based on unpolarized incident light. Considering that the linearly polarized laser is usually utilized as the light source in experiments, the perpendicularly polarized Rayleigh ratio $R_v$ can be deduced in accordance with Krishnan's relations at $90^\circ$ ~\cite{Krishnan1939},
\begin{equation}
	R = \frac{1+\delta}{2}R_v,  \label{Eq:R2Rv}
\end{equation}
where $v$ in the subscript represents the polarization direction of the incident light is perpendicular to the scattering plane. Then the expression for the Rayleigh scattering length can be rewritten as:
\begin{equation}
	L_\mathrm{Ray} = \left[ \frac{8\pi}{3}R_v\frac{2+\delta}{2}\right]^{-1}. \label{Eq:finalLRay}
\end{equation}
As the knowledge on the depolarization ratio $\delta$ can be found in Ref.~\cite{Liu2015,Wurm:2010ad,Zhou:2015fwa}, an average value $\delta=0.31\pm0.02$ from all the measurement results is adopted. Therefore, the purpose for this experiment is to precisely determine the perpendicularly polarized Rayleigh ratio $R_v$. Furthermore, one important property for Rayleigh scattering is the linear response, which makes it possible to validate measurement results in a rather large light intensity range.

\section{Experimental details}\label{Sec:Exp}
\subsection{Experimental setup} \label{SubSec:setup}
\begin{figure*}
	\centering
	\includegraphics[width=0.95\textwidth]{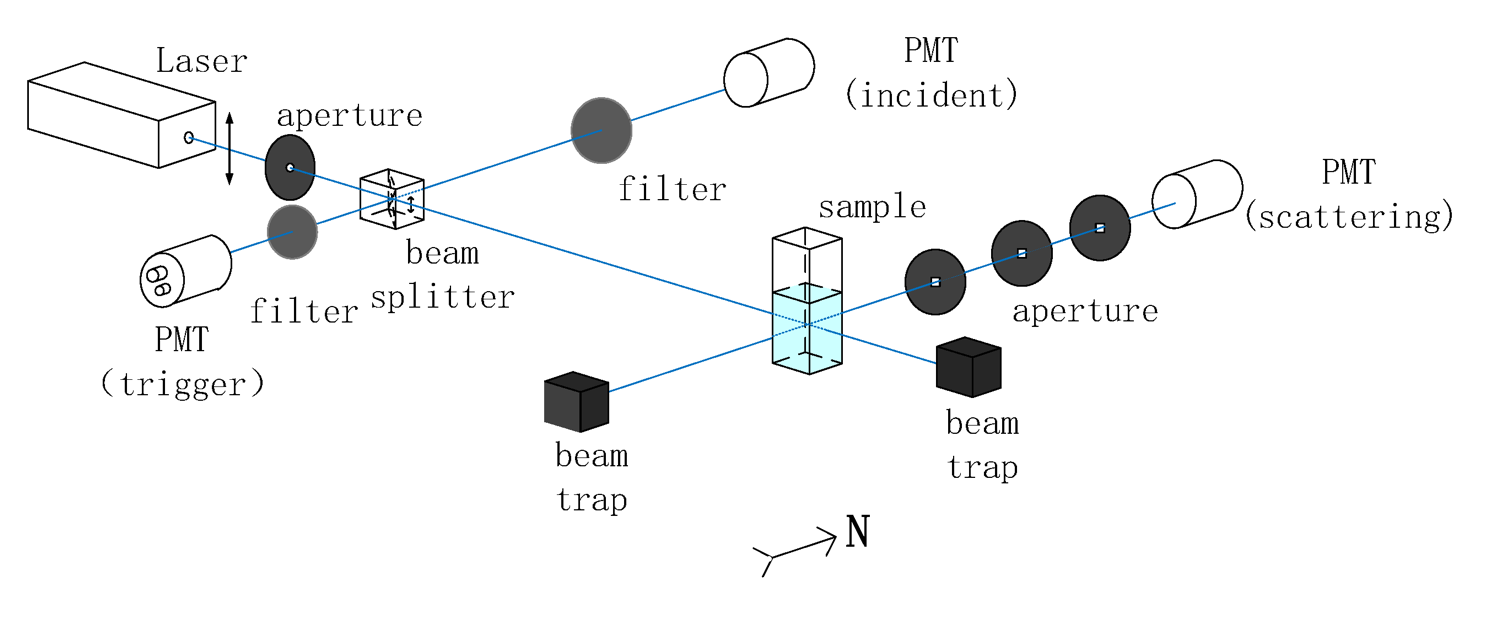}
	\caption{Sketch diagram of the experimental setup.}
	\label{Fig:setup}
\end{figure*}

The experiment setup to measure the polarized Rayleigh ratio $R_v$ for LAB is schematically shown in Fig.~\ref{Fig:setup}. An important issue was background suppression for such an experiment. To achieve this goal, the whole system was positioned in a dark room, meanwhile compact optical modules like tubes and cages were utilized for light path encapsulation. The temperature was kept at $\SI{21}{\celsius}$ stably. The light source, a Pico-Quant LDH pulsed diode, provided perpendicularly polarized light beams at two available wavelength peaks, $\SI[separate-uncertainty = true]{405.3\pm0.8}{nm}$ and $\SI[separate-uncertainty = true]{432.3\pm0.3}{nm}$ with adjustable output power up to $\SI{10}{mW}$ at $\SI{40}{MHz}$ repetition rate. A circular aperture with $\SI{0.1}{mm}$ diameter after the laser source was used for beam collimation and scattering volume definition. The incident beam was split by a beam splitter (Thorlabs CCM1-BS013) with one beam in the original direction (front), and the other beam was reflected into $90^{\circ}$ light path (left) used for intensity fluctuation monitor. LAB samples were held in a quartz cubic cuvette with $\SI{1}{cm}$ light path. The refractive index of the cuvette was close to that of LAB. The scattering measurement path was constructed at $90^{\circ}$ direction. Three square apertures at the scattering path were used to determine scattering volume and also to suppress stray light. Two beam traps were positioned at the other two sides of the cuvette also for backgrounds reduction. 

Because Rayleigh scattering of LAB is relatively weak, the intensity of the scattered light was expected to be at the single photon level. Therefore, the detection strategy in this experiment was photon counting method with photomultiplier tubes (PMTs). The scattered photons were counted by a Hamamatsu R1828-01 PMT, and the fluctuations of incident light were monitored by a Hamamatsu R2038 PMT. Because the intensity of the incident light beam exceeded PMT's dynamic range, $13$ absorptive attenuators (Thorlabs absorptive neutral density filters, maximum tolerable power $\SI{1}{W}$) were installed before the monitor PMT. Meanwhile, weak reflected light from the beam splitter to the opposite direction of the incident monitor path was detected by an ET 9841B PMT to provide triggers for data-taking. All the three PMTs were aligned into north-south orientation to avoid the effects from the geomagnetic field. 

PMT signals coincident with triggers were counted by a CAEN Mod. N1145 quad scaler and preset counter in this experiment. A forced delay data-taking window after the signal window was applied to estimate accidental coincidence rate (including dark noise, stray light effects). The most prominent advantages of this data-taking strategy were its easy handling data formats and high efficiency due to no waveform sampling. 

The basic measurement idea for the perpendicularly polarized Rayleigh ratio $R_v$ is to compare the light intensity of the scattered beam with that of the incident beam by the definition. So the explicit expression for the experimental determined $R_v$ can be written as:
\begin{equation}
	R_{v}=\frac{N/\mathrm{d}S}{N_{0}/\mathrm{d}S_0}\frac{r^2}{V}=\frac{N/\mathrm{d}\Omega}{N_{0}/\mathrm{d}S_0}\frac{1}{L\cdot \mathrm{d}S_0}=\frac{N/\mathrm{d}\Omega}{N_0L},
	\label{Eq:detailedRR1}
\end{equation}
where $N$ and $N_0$ refer to the real scattered and perpendicularly polarized incident photon numbers, $L$ is the distance from the scattering center to the detector and $\mathrm{d}\Omega$ is the scattering solid angle defined by apertures in the scattering light path. The furthest aperture from the PMT has height $H$ and width $W$ and so the solid angle $\mathrm{d}\Omega = \frac{W \cdot H}{L^2 n^2} (W, H \ll L)$. Here the differences between refractive indices of air and LAB bring a geometric correction term $\alpha(n)$, which will be discussed in Section \ref{Sec:Uncer}. Furthermore, by introducing the attenuation coefficient of attenuators $\epsilon_\mathrm{att}$, the split ratio of the beam splitter $\epsilon_\mathrm{BS}$ and the relative detection efficiency (DE) of the two PMTs $\epsilon_\mathrm{PMT}^r$, the Rayleigh ratio can be rewritten as:
\begin{equation}
	R_v=\frac{\tilde{N}}{\tilde{N_0}}\epsilon_\mathrm{att}\epsilon_\mathrm{BS}\epsilon_\mathrm{PMT}^r\frac{r^2\alpha(n)}{LWH},
	\label{Eq:detailRR2}
\end{equation}
where $\tilde{N}$ and $\tilde{N_0}$ represent the detected scattered and incident photon numbers involved all instrumental effects. Moreover, with rotatable polarizers installed in the scattering light path, the depolarization ratio $\delta$ is measurable based on the current experimental setup.

\subsection{Calibration} \label{SubSec:Calib}
From Eq.~\eqref{Eq:detailRR2}, the calibration for the attenuation coefficient $\epsilon_\mathrm{att}$, the split ratio $\epsilon_\mathrm{BS}$ and the relative DE $\epsilon_\mathrm{PMT}^r$ at both $\SI{405}{nm}$ and $\SI{432}{nm}$ is essential for Rayleigh ratio measurements. All calibration results are visualized only for $\SI{432}{nm}$ as examples in the following sections.

\subsubsection{Attenuation coefficient of attenuators} \label{SubSubSec:att}
\begin{figure*}
	\centering
	\includegraphics[width=0.95\textwidth]{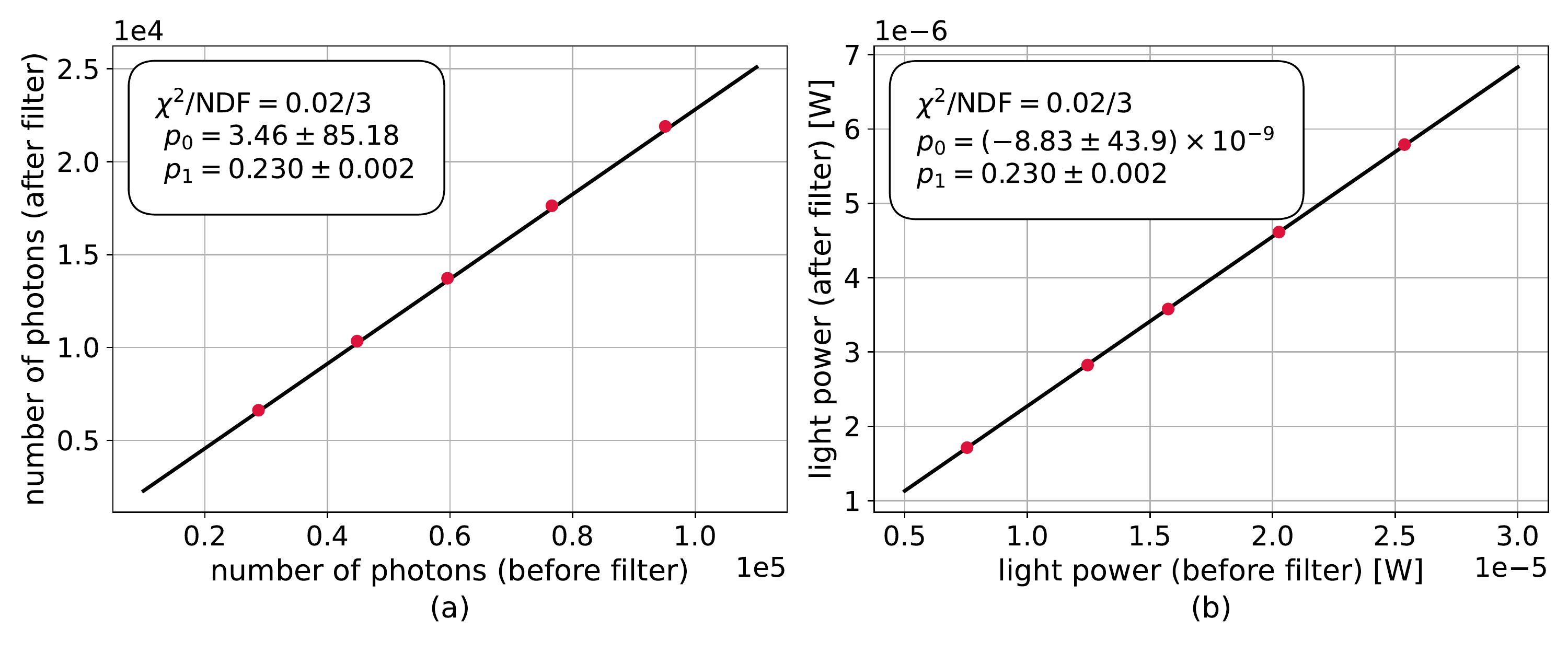}
	\caption{Attenuation coefficient calibration of the $3$rd attenuator at $\SI{432}{nm}$ with PMTs (a) and the power meter (b). Linearity of the attenuation system was validated in a large light intensity scale. The attenuation coefficient was calibrated by a linear function fitting ($y=p_0+p_1\times x$) of light intensity before and after the attenuator, and the two methods give consistent results within $1\sigma$ statistical uncertainties ($p_1 = \epsilon_\mathrm{att}^3$ in legends). }
	\label{Fig:OneFilter_PMT+PM}
\end{figure*}

Around $10^9$ times intensity decrease after attenuation made the direct calibration of $\epsilon_\mathrm{att}$ impossible. Thus attenuation coefficient for each single attenuator $\epsilon_\mathrm{att}^i$ was calibrated separately and the total attenuation coefficient was given by
\begin{equation}
	\epsilon_\mathrm{att} = \prod_{i=1}^{13}{\epsilon_\mathrm{att}^i}.
	\label{Eq:attcoeff}
\end{equation}
Because the working environments for all attenuators span a large light intensity range, it's necessary to ensure that the linear response was satisfied. The calibration under high intensity light with a power meter NEWPORT 2936-R validated the good linearity for this system. Results of one attenuator are shown in Fig.~\ref{Fig:OneFilter_PMT+PM} (b), and consistency was found within $1\sigma$ uncertainty compared with the results calibrated with PMTs under weak light conditions in Fig.~\ref{Fig:OneFilter_PMT+PM} (a).

\begin{figure*}
	\centering
	\includegraphics[width=0.95\textwidth]{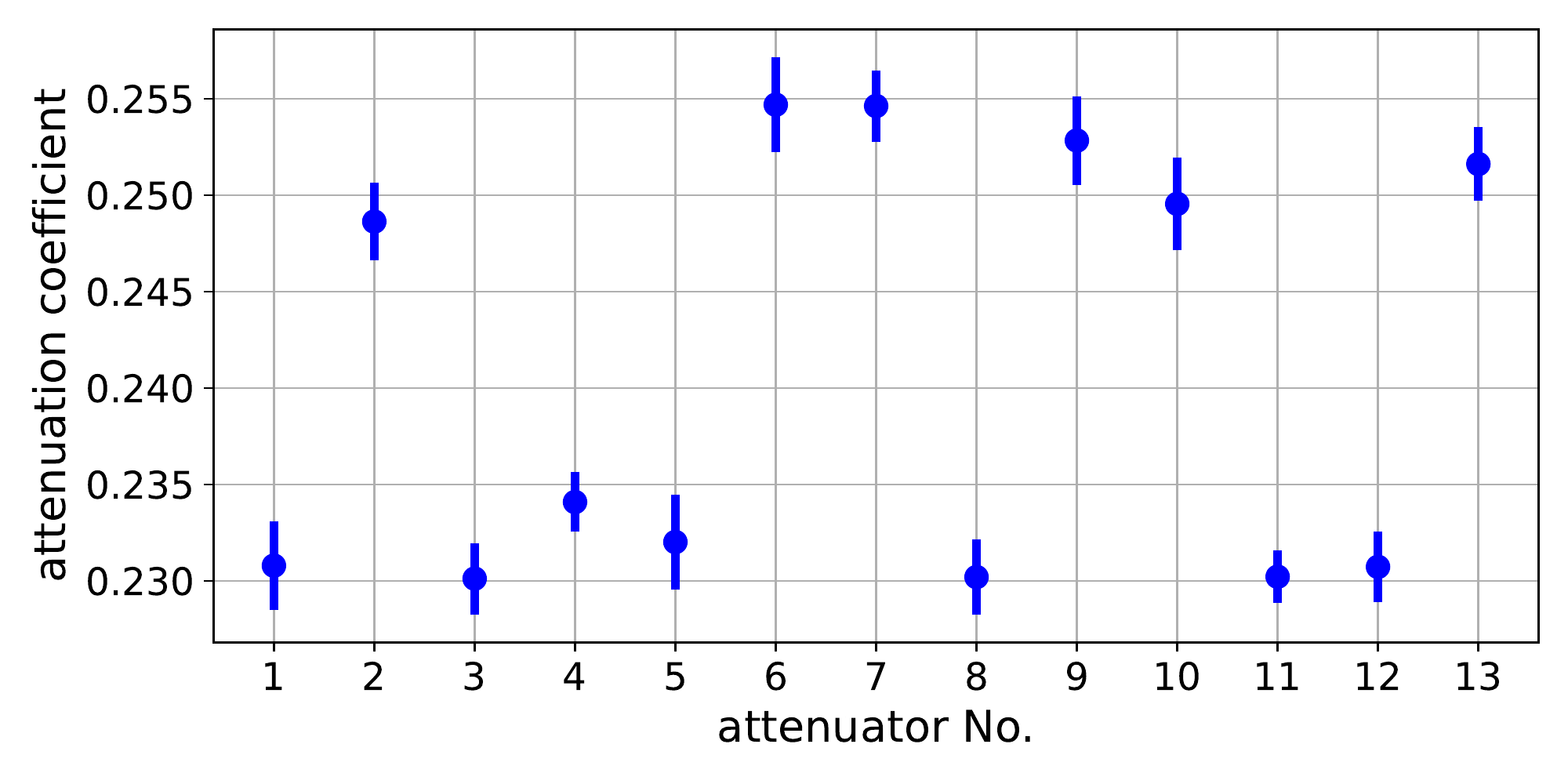}
	\caption{Attenuation coefficient calibration of $13$ attenuators at $\SI{432}{nm}$ with PMTs.}
	\label{Fig:att_calib}
\end{figure*}

The calibration results with PMTs for all $13$ attenuators at $\SI{432}{nm}$ are shown in Fig.~\ref{Fig:att_calib}. According to Eq.~\eqref{Eq:attcoeff}, the calculation values and uncertainties of $\epsilon_\mathrm{att}$ at both $\SI{405}{nm}$ and $\SI{432}{nm}$ are listed in Table \ref{tab:calib}. 

\subsubsection{Split ratio of beam splitter}\label{SubSubSec:split}
\begin{figure*}
	\centering
    \includegraphics[width=0.7\textwidth]{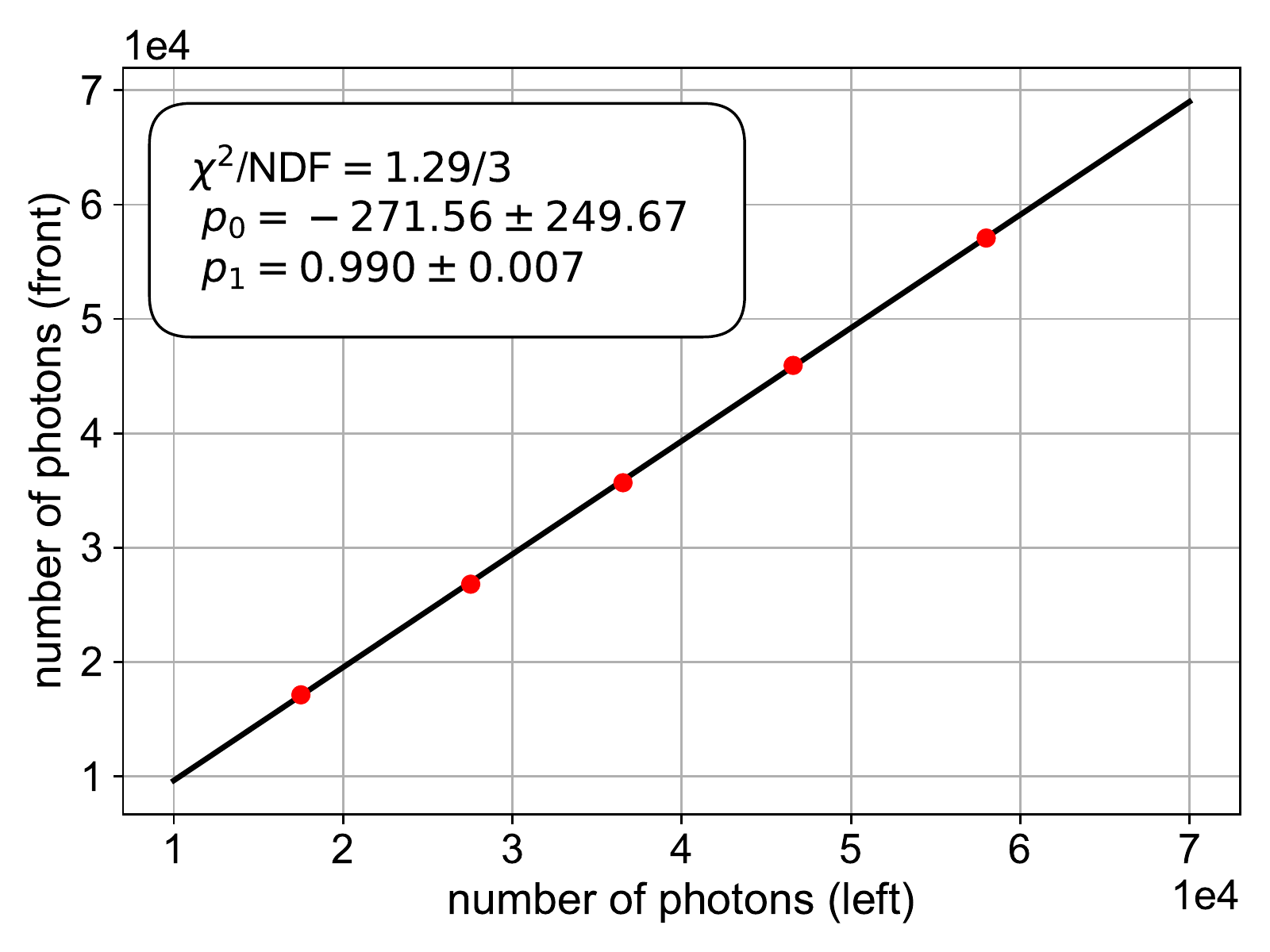}
	\caption{Calibration of split ratio at $\SI{432}{nm}$ with PMTs. A linear fitting was applied for the intensity of two split light beams ($p_1 = \epsilon_\mathrm{BS}$ in the legend).}
	\label{Fig:splitter}
\end{figure*}

The split ratio was determined by the relative ratio of the light intensity at the scattering light path and the incident monitor path measured by a same PMT under unchanged working conditions. Besides the single photon counting method, the ratio of more intense light, with power in the order of magnitude microwatt, was also calibrated by the power meter to provide linearity validation for the beam splitter. The final results of the split ratio are shown in Fig.~\ref{Fig:splitter} and listed in Table.~\ref{tab:calib}.

\subsubsection{Relative detection efficiency of PMTs}\label{SubSubSec:DE}
\begin{figure*}
	\centering
    \includegraphics[width=10cm]{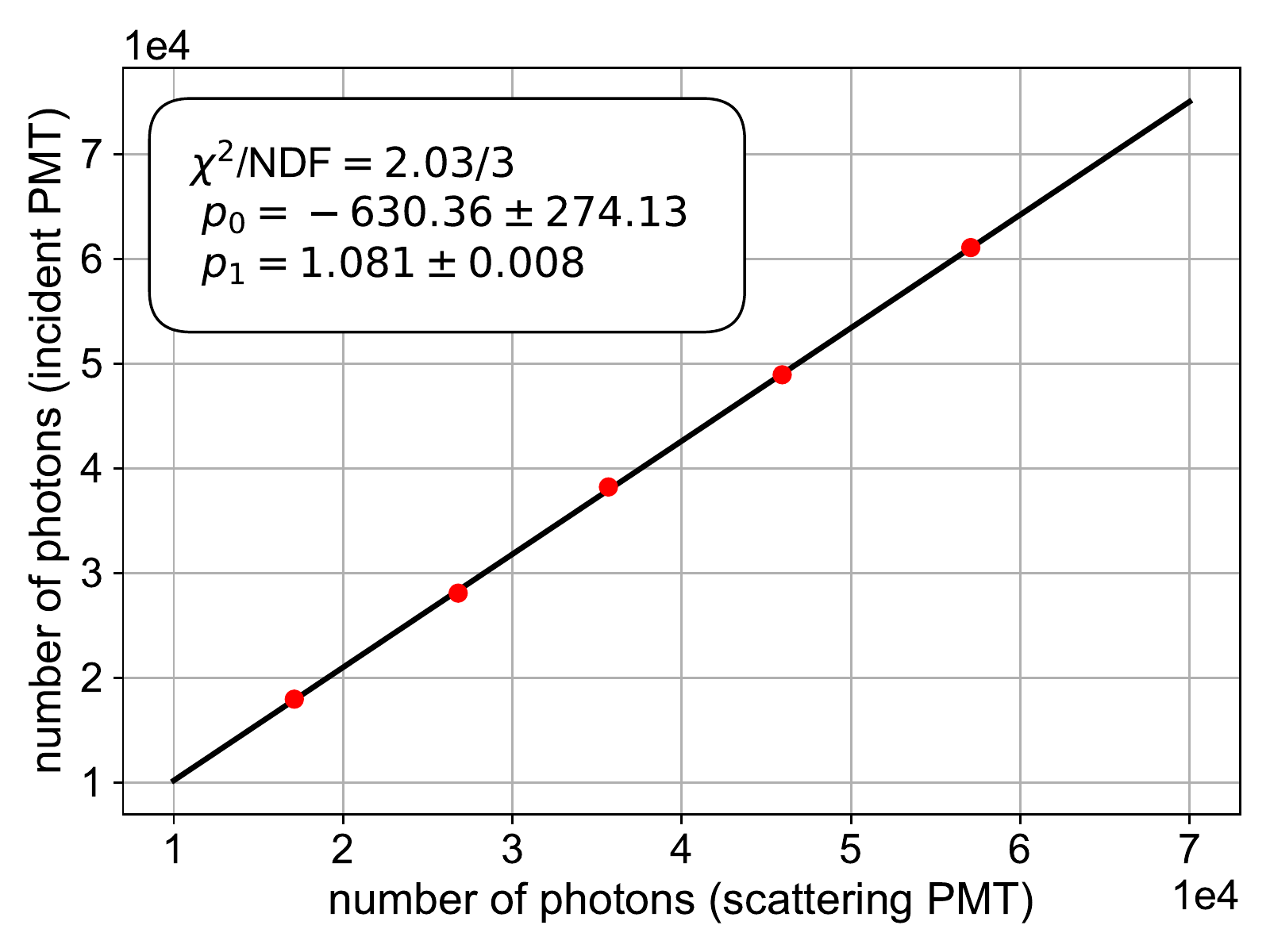}
	\caption{ Calibration of relative detection efficiency at $\SI{432}{nm}$. A linear fitting was applied for the detected photon numbers with two PMTs ($p_1 = \epsilon_\mathrm{PMT}^r$ in the legend)}
	\label{Fig:relaQE}
\end{figure*}

During the measurements, photons were collected at the scattering path and the incident monitor path simultaneously with two different PMTs. Thus, it is necessary to correct the relative detection efficiency. Wavelength dependency of DE was not concerned as Rayleigh scattering is an approximately elastic process. The calibration strategy was to measure photon numbers with the incident PMT and the scattering PMT placed at the same position in succession. The relative DE was obtained as the ratio of detected photon numbers. The calibration results are shown in Fig.~\ref{Fig:relaQE} and also listed in Table~\ref{tab:calib}.

\begin{table}[!htbp]
\centering
\begin{tabular}{ccc}
\toprule Wavelength (nm) & $\SI{405}{nm}$ & $\SI{432}{nm}$ \\
\hline
Attenuation coefficient ($\epsilon_\mathrm{att})$ & $(2.900\pm0.113)\times10^{-9}$ & $(9.024\pm0.280)\times10^{-9}$ \\
Split ratio ($\epsilon_\mathrm{BS})$  & $1.141\pm0.024$ & $0.990\pm0.007$  \\
Relative DE ($\epsilon_\mathrm{PMT}^r)$ & $1.227\pm0.019$ & $1.081\pm0.008$  \\
\hline
\end{tabular}
\caption{Calibration results of the attenuation coefficient $\epsilon_\mathrm{att}$, the split ratio $\epsilon_\mathrm{BS}$ and the relative detection efficiency $\epsilon_\mathrm{PMT}^r$ at $\SI{405}{nm}$ and $\SI{432}{nm}$.}\label{tab:calib}
\end{table}

\section{Systematic uncertainties}
\label{Sec:Uncer}
The systematic uncertainties of this experiment have been discussed from following aspects:

The wavelength distributions of the light source were fitted by gaussian functions and the peak positions were at $\SI{405.3\pm0.8}{nm}$ and $\SI{432\pm0.3}{nm}$. Thus the uncertainty from this contribution is negligible. Moreover, the intensity fluctuation of the laser source was cancelled by the simultaneous incident light monitor.

The calibration absolute uncertainties for the attenuation coefficient $\epsilon_\mathrm{att}$, the split ratio $\epsilon_\mathrm{BS}$ and the relative detection efficiency $\epsilon_\mathrm{PMT}^r$ are given in Table~\ref{tab:calib} and included in the total uncertainty calculation.

The whole system was compact and modularized which kept the backgrounds at a low level. Uncertainties from stray light during measurements were estimated to be $4.0\%$ by scattering PMT measurements with an empty cuvette before LAB was filled.

The geometric correction term $\alpha(n)$ mentioned in Section~\ref{SubSec:setup} is a correction on the solid angle of view for the scattering PMT. For ideal point light source $\alpha_\mathrm{point}(n) = n^2$, where $n$ is refractive index of LAB and also the cuvette. However, this term requires corrections under different circumstances according to the experimental layouts such as the length of scattering light path and the geometry of the apertures~\cite{Carr1950}. Instead of direct calculations of the correction, a Geant4 based~\cite{Geant42003} optical simulation was performed for specific studies. The experimental layout was applied in the geometry simulation, and a line light source was placed in the cuvette to mimic the scattering light path. A small deviation around $4\%$ has been observed between $\alpha_\mathrm{point}(n)$ and real $\alpha(n)$ from simulation results. In the final analysis, $\alpha(n) = n^2$ was applied while the $4\%$ level deviation was considered as the systematic uncertainty for this correction. Moreover, the effects of the bad alignment of the three apertures were studied with simulation. The uncertainty introduced by this effect was around $0.2\%$ and ignored in the calculation.

All systematic uncertainties discussed above are summarized in Table~\ref{tab:error}. The total systematic uncertainties are estimated as $7.4\%$ at $\SI{405}{nm}$ and $6.6\%$ at $\SI{432}{nm}$.
\begin{table}[!htbp]
\centering
\begin{tabular}{ccc}
\toprule Correction terms &  $\SI{405}{nm}$ & $\SI{432}{nm}$ \\
\hline
Attenuation coefficient  &  $3.9\%$ & $3.1\%$\\
Split ratio  & $2.1\%$ & $0.7\%$\\
Relative DE & $1.5\%$ & $0.7\%$ \\
Depolarization ratio  & $0.9\%$ & $0.9\%$ \\
$\alpha(n)$ & $4.0\%$ & $4.0\%$  \\
Stray light & $4.0\%$ & $4.0\%$  \\
\hline\hline
Total & $7.4\%$ & $6.6\%$  \\
\hline
\end{tabular}
\caption{Systematic uncertainties.}
\label{tab:error}
\end{table}

\section{Results and discussion}\label{Sec:Results}
\begin{figure*}
	\centering
    \includegraphics[width=0.95\textwidth]{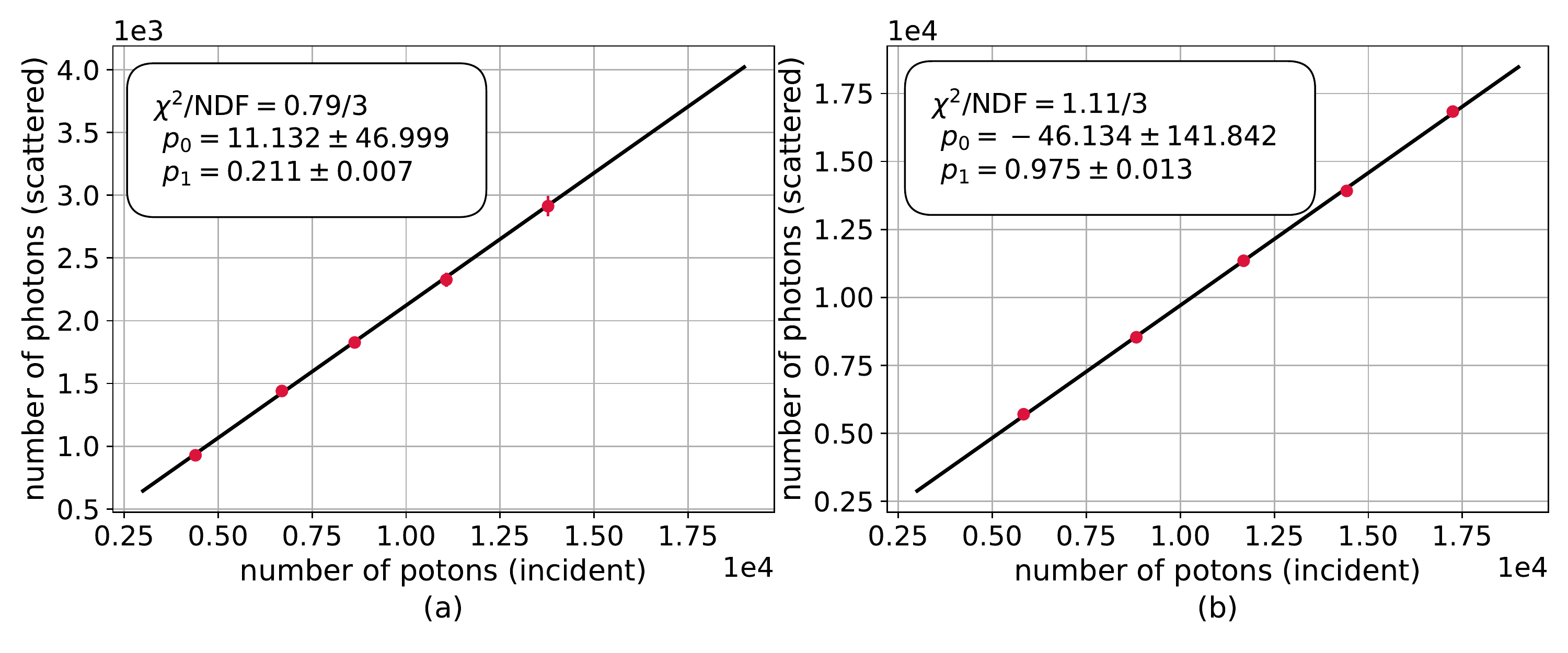}
	\caption{Measurements of scattered light intensity with respect to the incident light intensity for at both $\SI{432}{nm}$ (a) and $\SI{405}{nm}$ (b). Linear functions were applied for the intensity ratio fitting. And $p_1$ in legends refer to the values of ${\tilde{N}}/{\tilde{N_0}}$ required in Eq.~\eqref{Eq:detailRR2}.}
	\label{Fig:photonRatio}
\end{figure*}

The measurements of photon numbers of the scattered light and the incident monitor light ${\tilde{N}}/{\tilde{N_0}}$ at two wavelength are shown in Fig.~\ref{Fig:photonRatio}.
Because of the linear response of Rayleigh scattering process and the linearity calibration of all optical components in Section~\ref{SubSec:Calib}, linear fitting is adopted in the analysis and performs well as the expectation. With the usage of calibration results for all correction terms, the perpendicularly polarized Rayleigh ratio $R_v$ is measured as $\SI{4.52+-0.28e-6}{m^{-1}.sr^{-1}}$ at $\SI{405}{nm}$ and $(3.82\pm 0.24)\times 10^{-6}$ m$^{-1}\cdot$ sr$^{-1}$ at  $\SI{432}{nm}$ according to Eq.~\eqref{Eq:detailRR2}. Thus based on the current knowledge on the depolarization ratio $\delta$, the corresponding values for the Rayleigh scattering length is $L_\mathrm{Ray} = 22.9\pm 0.3(\mathrm{stat.})\pm 1.7(\mathrm{sys.}) {\rm m}$ at $\SI{405}{nm}$ and $L_\mathrm{Ray}= 27.0\pm 0.9(\mathrm{stat.})\pm 1.8(\mathrm{sys.}){\rm m}$ at $\SI{432}{nm}$.
\begin{figure*}
	\centering
    \includegraphics[width=10cm]{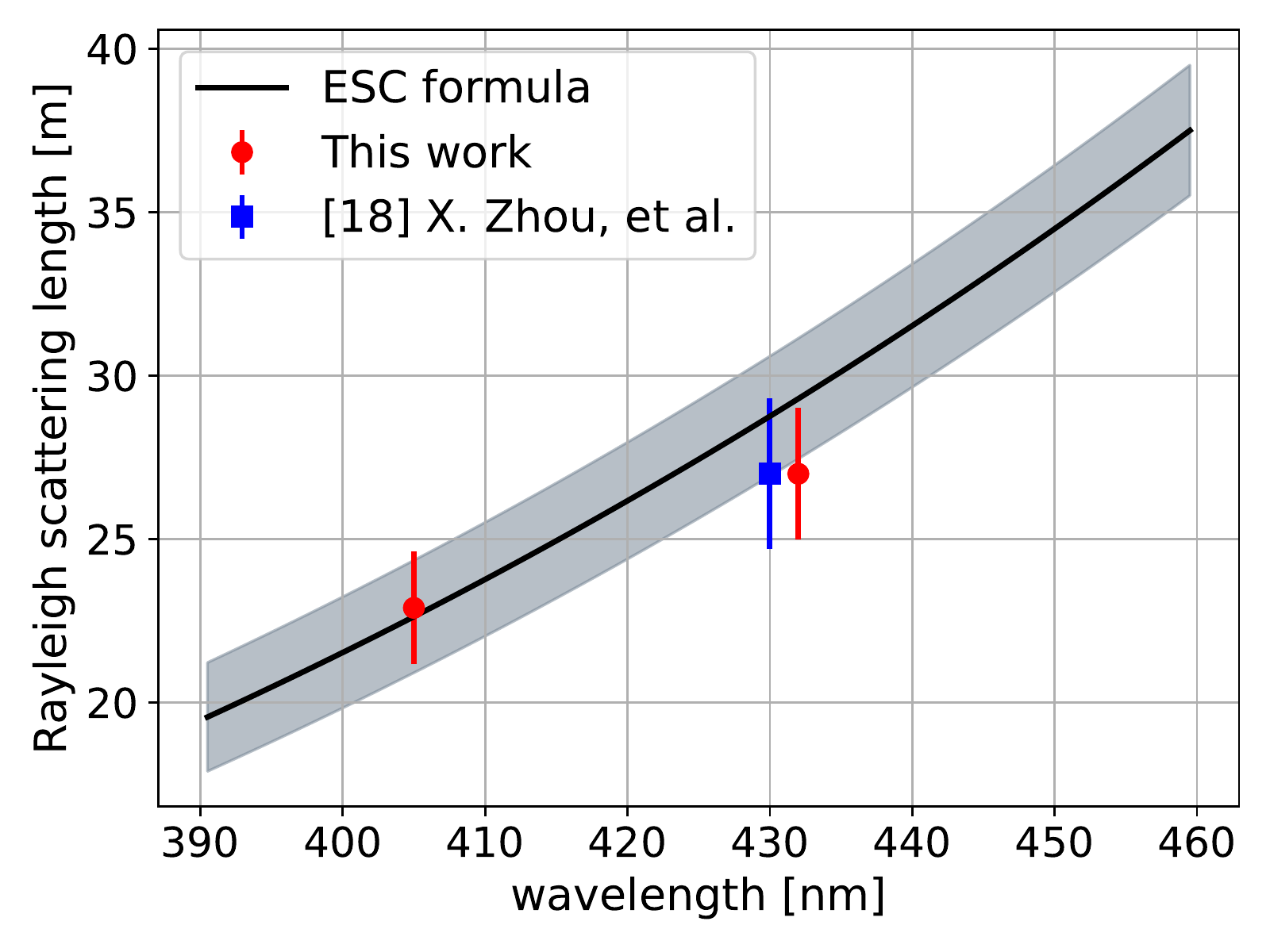}
	\caption{Illustration of the Rayleigh scattering length of LAB in theoretical calculation and experiment measurements. The black line represents the ESC formula calculation and the shadow band represents $1\sigma$ area. Red markers are the measurement values in this work. The blue marker is the previous measurement result.}
	\label{Fig:ESC}
\end{figure*}

As discussed in previous studies, the Rayleigh scattering length of liquids can be described empirically by the ESC formula. It is shown that the Eykman equation performs well for organic liquids~\cite{Eykman1895}. And the ESC formula is expressed as:
\begin{equation}
	L_{Ray} = \left[ \frac{8\pi^3}{3\lambda^4} \left[ \frac{(n^2-1)(2n^2+0.8n)}{n^2+0.8 n+1}\right]^2kT\beta_T \frac{6+3\delta}{6-7\delta} \right]^{-1},
	\label{Eq:ESC}
\end{equation}
where $\lambda$ is the wavelength of the scattered light, $n$ is the refractive index, $k$ is the Boltzmann constant, $T$ is the absolute temperature, $\beta_T$ is the isothermal compressibility. Using the measured values of LAB properties in Ref.~\cite{Zhou:2014wra}, the measured Rayleigh scattering length can be compared with theoretical prediction in Fig.~\ref{Fig:ESC} and consistency can be found within $1\sigma$ uncertainty. Also the measurement values are close to the results from previous experiments with slightly smaller uncertainties.

In the present work, we have directly measured the Rayleigh scattering length and consistent results with the previous measurements have been obtained. Several advantages for the experimental design make it possible to be generally used in Rayleigh scattering measurements. An important merit is the low time cost benefiting fromv the single angular data-taking and photon counting strategy. Stable environment temperature is easier to realize in a relatively shorter time interval. Besides, the whole experimental setup is designed to be compact and modularized, which facilitates the background control and commercial usage. From the perspective of theory, the model independent method provides opportunities to test the theoretical model of Rayleigh scattering, for example, the performance of Eykman equation approximation. Moreover, measurements of the Rayleigh scattering length and the absorption length are valuable for Monte Carlo studies of LS detectors, beneficial for both detector designs and algorithm developments. 

\section*{Acknowledgments}
This work has been supported by the National Natural Science Foundation of China (Grant No.12175241) and the Major Program of the National Natural Science Foundation of China (Grant No.11390381).





\end{document}